# SAM: Support Vector Machine Based Active Queue Management

MUHAMMAD SALEH SHAH*, ASIM IMDAD WAGAN**, AND MUKHTIAR ALI UNAR***

## ABSTRACT

Recent years have seen an increasing interest in the design of AQM (Active Queue Management) controllers. The purpose of these controllers is to manage the network congestion under varying loads, link delays and bandwidth. In this paper, a new AQM controller is proposed which is trained by using the SVM (Support Vector Machine) with the RBF (Radial Basis Function) kernal. The proposed controller is called the support vector based AQM (SAM) controller. The performance of the proposed controller has been compared with three conventional AQM controllers, namely the Random Early Detection, Blue and Proportional Plus Integral Controller. The preliminary simulation studies show that the performance of the proposed controller is comparable to the conventional controllers. However, the proposed controller is more efficient in controlling the queue size than the conventional controllers.

## 1.     INTRODUCTION

Network congestion, its control and avoidance is very hot area for research. Although the bandwidth has increased a lot since last decade but still the problem of congestion occurs due to the changing nature of the traffic. A major part of the network management is the design of the new queuing techniques in the router. One such technique is known as the AQM which tries to reduce the congestion problem by intelligently managing the queue. AQM plays a major role in network congestion by controlling the network congestion and actively managing the queue. Several different algorithms are introduced in the AQM literature such as RED (Random Early Detection) [1], Blue [2] and PI (Proportional Integral) [3]. All algorithms have their own design and different parameters with some advantages and problems. The research is being carried out to get more and more stable controller.

SVMs are robust classifiers and can be used in variety of fields where classification is required. SVMs can predict the unknown data. This paper introduces SAM controller that actively manages the queue using SVM algorithm.

The remaining paper is organized as follows. Section 2 provides the back ground of network congestion and avoidance algorithms and their classification like TCP

\*          Ph.D. Scholar, Institute of Information & Communication Technologies, Mehran University of Engineering & Technology, Jamshoro.
\*\*         Professor, Department of Computer Engineering, DHA Suffa University, Karachi.
\*\*         Professor, Department of Computer Systems Engineering, Mehran University of Engineering & Technology, Jamshoro.

(Transmission Control Protocol) window and AQM. It gives the review of existing controllers, their design concepts and problems associated with them. At the end, this section provides the SVM overview and its need in network congestion control. Section 3 describes SAM and its basic design. Section 4 compares results of SAM with RED, Blue and PI. Finally, Section 5 concludes with prospects for future work.

## 2. PREVIOUS WORK

It is very essential to introduce robust algorithms to control and avoid the network congestion. It is also important to analyze the performance of the controllers under various network conditions. These algorithms are largely categorized into two. One is the algorithms that are applied in TCP like TCP Reno and TCP Vegas, for adjusting the transmission rate according to congestion feedback. Other are the AQM algorithms that are applied at link nodes, such as RED, Blue and PI for deciding which arriving packets to drop under network overload conditions.

TCP is connection oriented and reliable protocol. TCP is the dominating protocol throughout the world due its features since last two decades. It identifies the congestion only after a packet has been dropped [4]. It uses TCP window to avoid the network congestion with slow start and gradual rise in window size if acknowledgment is received and vice versa [5].

AQM is considered as an effective way for congestion control [6]. It effectively achieves a balance between link utilization and delay [7]. It controls the queue before queue becomes full and packet drops occurs. AQM maintains the small queues and reduces the delays. AQM has natural capacity to absorb packet bursts without any drops. During overflow conditions more packets drop without AQM [8].

The major objectives of AQM are to maintain stable and fair queues [9]. RED, FRED, Blue, SFB and CHOKe were compared and performance of RED and Blue was found better than others in link utilization and others maintained fairness [10].

RED algorithm was introduced by Floyd and Jacobson in 1993 [1]. It is widely used AQM controller. RED maintains two queue thresholds, such as minth and maxth. No packet will be dropped when queue is less then minth, but, if it exceeds maxth the all incoming packets will be dropped. The RED controller is composed of two major parts. First is the average queue size estimation and other is the decision function to drop the packets. The marking probability will be maintained if the queue is between minimum and maximum [8].

RED has some major issues related with its design. If it is designed aggressively it will show extremely fast response but low margin for stability. If it is designed to be more stable it has very slow response time. RED creates a relationship between queue length and loss probability which results in steady state errors [3]. RED causes oscillations and instability due to the parameter variations [11].

Blue algorithm was introduced by Feng, et. al. [2]. The Blue AQM controller manages queue based link ideal time and packet loss instead of queue lengths. Blue maintains a single Pm (marking probability) to mark/drop packets. Blue decreases its Pm when link is idle and increases Pm when it finds continuously packets drop due to buffer overflow [12].

Problems associated with Blue are large queue size cannot absorb TCP Bursts and packets have to wait (Increased Round Trip Time). Its performance also suffers with wide variation in network load [8].

The PI is well known controller that is widely utilized in the industry. It is highly robust and easy to understand. It results in good performance under instability. PI AQM controller decides for packet dropping based on its dropping probability. This dropping probability is calculated from old probability, current difference of queue length with reference queue and old difference of queue length with reference queue.

Problems associated with PI are that, it is only suitable for linear and deterministic systems and signals. Internet/Network congestion is non linear and stochastic in nature. The performance of PI therefore cannot be optimal. The linearization introduces modeling errors. Average queue length increases as number of senders increases. It mainly depends on reference queue [8].

SVMs are one of the successful classifiers in industry and academia. Due to their performance and accuracy to solve the complex problems they are used extensively. There are two major categories of SVMs, Linear and Non-Linear. Their models are robust to parameter deviation and capable to generalize the unseen data [13,14]. They are also well recognized in continuous adaptive learning. These features are extremely necessary in network environments [13].

## 3. SAM: SVM BASED AQM

The idea behind this controller is to actively manage the queue using machine learning algorithm. SVMs are machine learning algorithms and proven classifiers used in various fields. SAM uses SVM algorithm and its design is flexible to use different SVM kernels. This research contains only RBF kernel responses or simulation results of proposed controller with comparison to other existing controllers (RED, Blue and PI). The proposed controller uses buffer utilization (queue size) patterns for packet drop decisions. It takes the buffer utilization pattern at the time of last five packets' arrival.

This research also includes a method to generate the feature vector from the possible data set of buffer utilization patterns. The SAM needs model of data set after training. The SVM model is generated through training on the selected data set. The decisions of proposed controller are directly dependent on the training and provided data set for it.

The proposed SVM based AQM controller is written in C++ and implemented on Network Simulator - 2 (NS-2) by adding the new AQM in NS-2's queue class. This controller also uses the SVM library during execution and training. It passes model file and the queue pattern to the SVM library. SVM library classifies or decides about packet to be dropped or en-queue.

## 4. RESULTS

This section contains the comparisons between the AQM controllers RED, Blue, PI and SAM 1 MB (Mega Bytes) bandwidth, 10ms (milliseconds) link delay and network load of 200 HTTP (Hyper Text Transfer Protocol) and 100

FTP (File Transfer Protocol) connections. Initially the response of each controller is separately discussed. These the overall performance of each controller is discussed on the basis of final results. The performance is evaluated in terms of throughput, network overload and round trip time. The scenario is simulated for 03 minutes or 180 seconds on NS-2. The general diagram of scenario is given in Fig. 1.

As shown in Fig. 2, packet arrival rate of RED is stable and low at start till 110 seconds then after it significantly increased and remains stable till the end. Packet arrival rate of Blue is marginally stable and lower than all other controllers. Packet arrival rate of PI is higher than others with gradual increase till 110 seconds, then it reduces approximately near to SAM. Packet arrival rate of SAM is most stable and significantly higher than Blue throughout the simulation. These results show the impact of controller's behavior on response of TCP. The prime responsibility of the controller is to increase throughput, decrease unnecessary network overload and reduce round trip time. Packet's arrival rate is shown in Fig. 2.

Packet departure rate of RED and PI is almost similar and is low till initial 110 seconds, then it significantly increase and remain stable and higher than the others till end of the simulation. Response of Blue remains lower than all other controllers. SAM shows best response in the starting and remains most stable throughout the simulation. These results show the behavior of each controller with respect to packet departure rate or throughput and the stability of each controller. Controllers' response in terms of packet departure rate or throughput is plotted in Fig. 3.

Packet drop rate of RED is low with lot of oscillations till initial 110 seconds, then it significantly increases the drop rate and decrease the oscillations till the end of simulation. Response of Blue is the best, remains stable with lower drop rate than all other controllers. PI gradually increases its drop rate at the start and remains with highest drop rate till the end. This clearly describes the worst performance of PI in terms of network overload. SAM responses with long oscillations and the center of oscillations remain stable throughout the simulation. These results show the behavior of each controller with respect to packets' drop rate and their performance in terms of network overload. Packet drop rate is shown in Fig. 4.

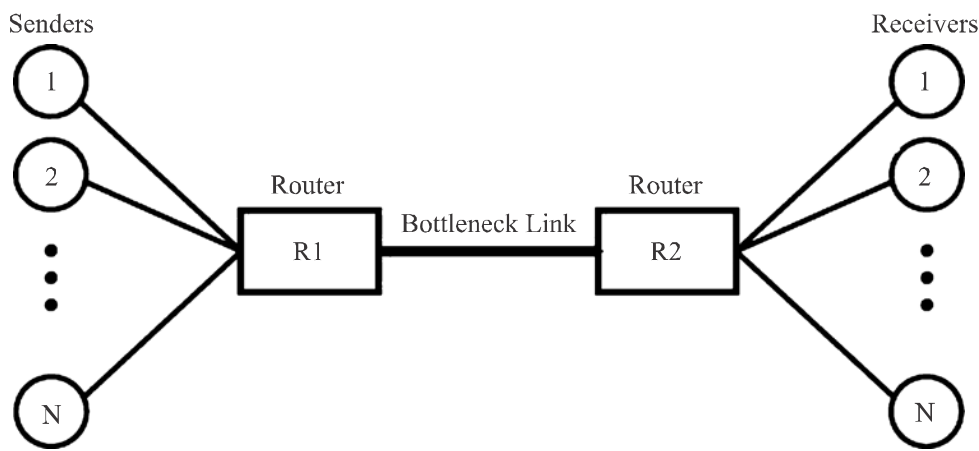

*FIG. 1.*

Queue size or buffer utilization of RED is high till initial 110 seconds, then after that it is significantly decreased till the end of the simulation. Response of Blue is the worst with lot of oscillations and high queue size than all other controllers. This causes increased round trip time. PI gradually decreases its drop rate till 110 seconds and remains marginally stable till the end. SAM's response is most stable and remains stable throughout the simulation. These results show the behavior of each controller with respect to queue size and their performance in terms of buffer utilization and round trip time. Queue size with respect to the time is plotted in Fig. 5.

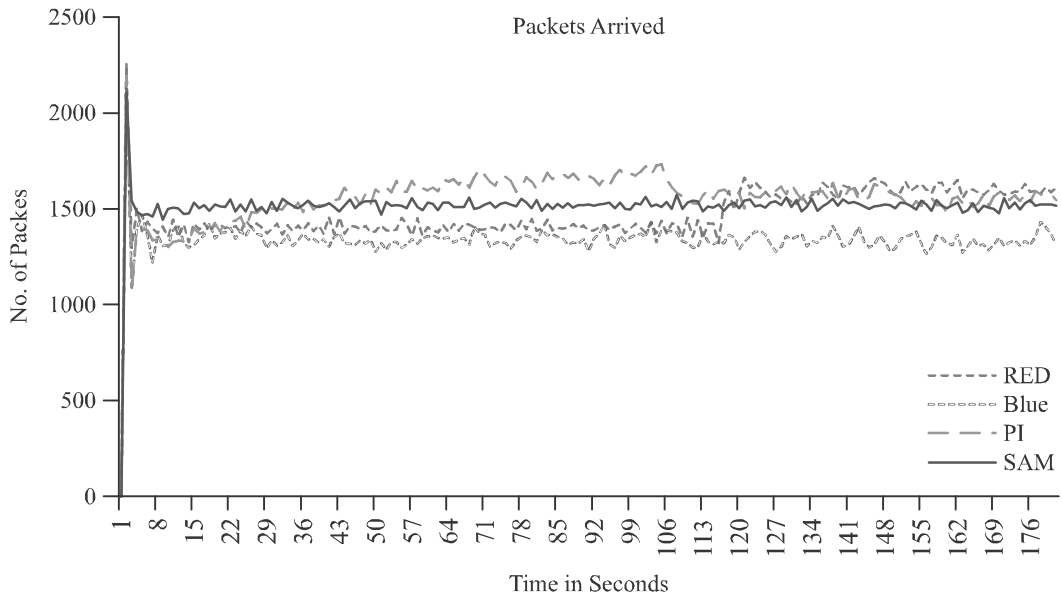

FIG.2.

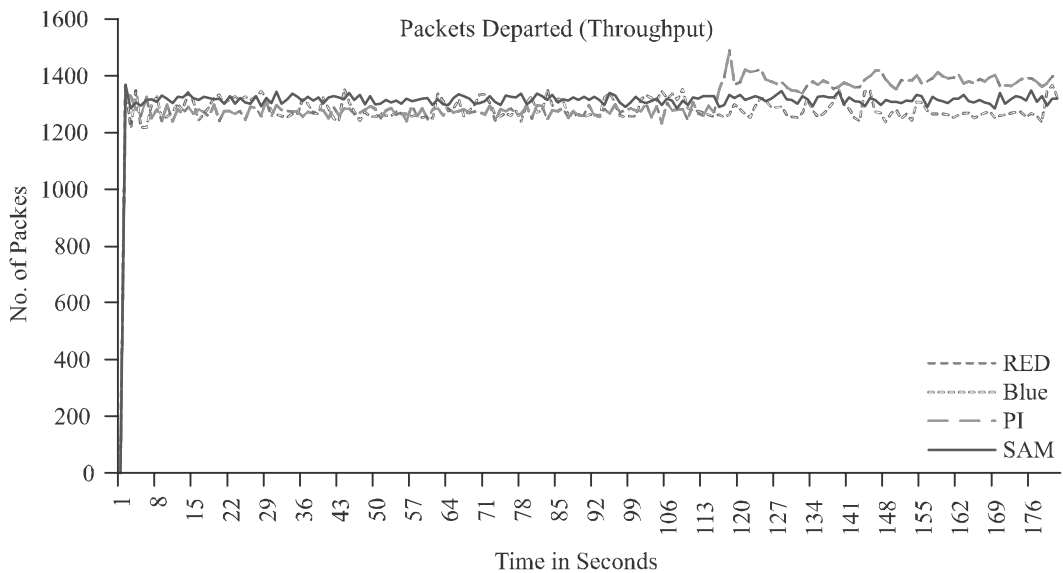

FIG .3.

The overall performance of the controllers has been summarized in Table 1. PI performs best and Blue performs worst in terms of throughput and Blue performs best and PI performs worst in terms of extra network overload. Similarly SAM performs better than RED in terms of throughput and RED performs better than SAM in terms of network overload. In broader view each controller has its own advantages and disadvantages.

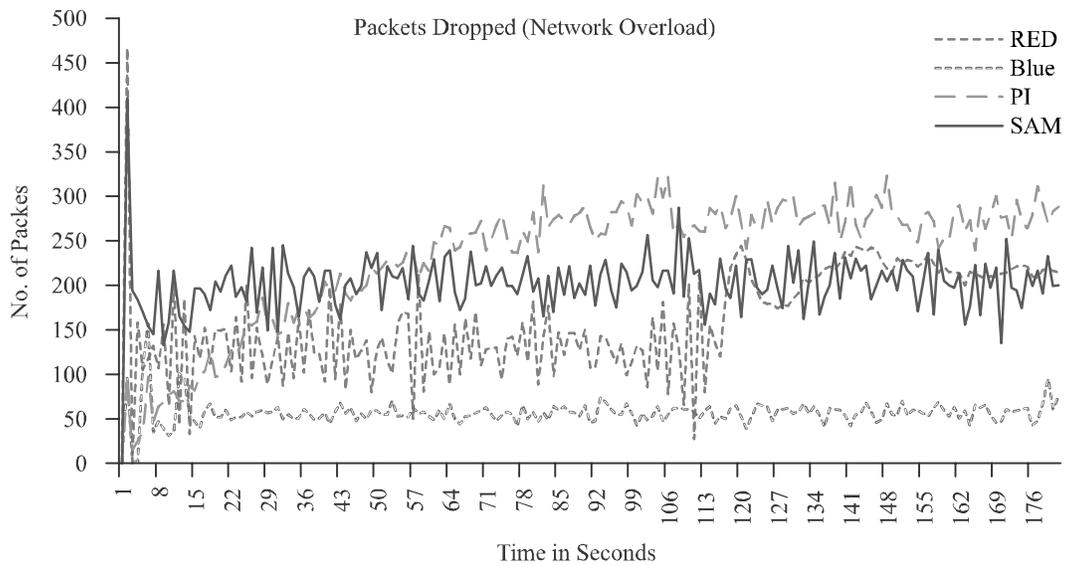

*FIG. 4.*

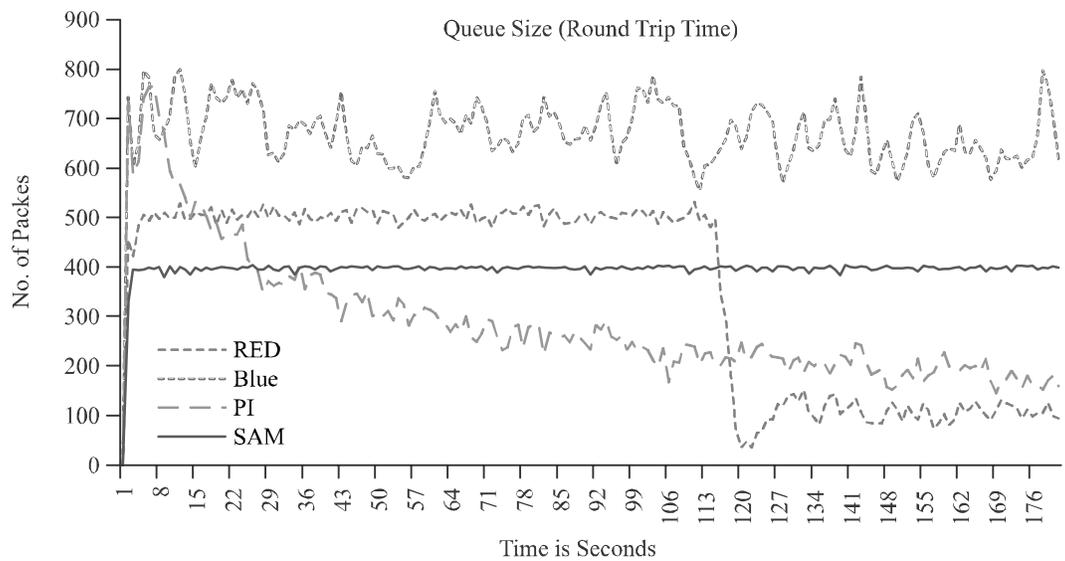

*FIG. 5.*

**TABLE 1.**

| No. | Controller | Total Arrivals | Total Departures | Total Drops | Average Queue Size |
|---|---|---|---|---|---|
| 1. | RED | 265,601 | 236,603 | 28,904 | 359 |
| 2. | Blue | 241,921 | 231,030 | 10,272 | 672 |
| 3. | PI | 281,642 | 236,603 | 41,951 | 291 |
| 4. | SAM | 273,815 | 236,901 | 36,516 | 397 |

## 5. CONCLUSIONS

This research introduces a new AQM controller based on SVM. This new controller is tested and simulated on the Network Simulator - 2 (NS-2). The proposed SVM based AQM controller is contrasted with other established AQM controllers such as RED, Blue and PI. The results show that SAM yields approximately similar performance as other AQM controllers but with an added advantage of controlled queue size. This has also been observed that SAM was stable throughout the scenario.

Further research can be carried out to optimize and enhance the performance of the SAM controller. In future, this AQM controller will be tested with better training models.

## ACKNOWLEDGEMENT

The authors acknowledged the suuport of the Department of Computer Systems Engineering, Mehran University of Engineering & Technology, Jamshoro, Pakistan, for conduct of this research work.